\documentclass{article}

\usepackage[margin=0.75in]{geometry}
\usepackage{booktabs} 
\usepackage[ruled]{algorithm2e}
\usepackage{amsmath,amssymb}
\usepackage{url,enumitem}
\usepackage{multirow,bigdelim}
\usepackage{booktabs}
\usepackage{array,graphicx}
\usepackage{float}
\usepackage{subfig}
\usepackage{wrapfig}
\usepackage{lscape}
\usepackage{rotating}
\usepackage{balance}
\usepackage{amsfonts}	
\usepackage[OT2,T1]{fontenc}
\usepackage[utf8]{inputenc} 
\usepackage{caption}
\graphicspath{{figs/}}
\usepackage{array} 
\usepackage{paralist} 
\usepackage{enumerate} 
\usepackage{verbatim} 
\usepackage{xparse}
\usepackage{listings}
\usepackage{color}  
\usepackage{hyperref} 
\usepackage{cleveref}
\usepackage{multirow}
\usepackage{multicol}
\usepackage{pdflscape}
\usepackage{rotating}
\usepackage{lipsum}
\usepackage{xspace}

\definecolor{dark-red}{rgb}{1,0.15,0.15}
\definecolor{dark-blue}{rgb}{0.15,0.15,1}
\hypersetup{
  colorlinks   = true, 
   urlcolor     = blue, 
   citecolor   = blue, 
}
\usepackage[textsize=small]{todonotes}

\usepackage{soul}
\setstcolor{red}
\setul{}{.2ex}

\usepackage{mathtools}

\usepackage[flushleft]{threeparttable}

\begin{document}
\title{Forward and Backward Feature Selection for \\ Query Performance Prediction}

\maketitle

{
\centering
\author{\textbf{Sébastien Déjean}}
\\IMT UMR5219 CNRS, UPS, Univ. de Toulouse, Toulouse, France.
\\sebastien.dejean@math.univ-toulouse.fr

\author{\vspace{0.5cm}\textbf{Radu Tudor Ionescu}}
\\University of Bucharest, Bucharest, Romania.
\\raducu.ionescu@gmail.com
 
\author{\vspace{0.5cm}\textbf{Josiane Mothe}}, 0000-0001-9273-2193
\\IRIT UMR5505 CNRS, INSPE, Univ. de Toulouse, Toulouse, France, F-31062.
\\Josiane.Mothe@irit.fr

\author{\vspace{0.5cm}\textbf{Md Zia Ullah}}
\\IRIT UMR5505 CNRS, Toulouse, France, F-31062.
\\mdzia.ullah@irit.fr
\\
}

\setlength{\abovedisplayskip}{3pt}
\setlength{\belowdisplayskip}{3pt}

\begin{abstract}
The goal of query performance prediction (QPP) is to automatically estimate the effectiveness of a search result for any given query, without relevance judgements. Post-retrieval features have been shown to be more effective for this task while being more expensive to compute than pre-retrieval features. Combining multiple post-retrieval features is even more effective, but state-of-the-art QPP methods are impossible to interpret because of the black-box nature of the employed machine learning models. However, interpretation is useful for understanding the predictive model and providing more answers about its behavior.
Moreover, combining many post-retrieval features is not applicable to real-world cases, since the query running time is of utter importance.
In this paper, we investigate a new framework for feature selection in which the trained model explains well the prediction. We introduce a step-wise (forward and backward) model selection approach where different subsets of query features are used to fit different
models from which the system selects the best one. 
We evaluate our approach on four TREC collections using standard QPP features. We also develop two QPP features to address the issue of query-drift in the query feedback setting. We found that:
(1) our model based on a limited number of selected features is as good as more complex models for QPP and better than non-selective models;
(2) our model is more efficient than complex models during inference time since it requires fewer features;
(3) the predictive model is readable and understandable; and
(4) one of our new QPP features is consistently selected across different collections, proving its usefulness.
\end{abstract}

%
%


\makeatletter
\def\url@foostyle{%
  \@ifundefined{selectfont}{\def\UrlFont{\sf}}{\def\UrlFont{\ttfamily}}}
\makeatother

\urlstyle{foo}

\section{Introduction}
\label{sec:Intro}

In information retrieval (IR), query performance prediction (QPP) aims at automatically predicting the effectiveness of a system for a given query, without relevance judgments. QPP is useful to inform an IR system whether a query is difficult or not, allowing the system to process it differently. For example, in case of a difficult query, the system could either apply a specific automatic query reformulation or engage in an interactive session with the user in order to provide a better answer \cite{chifu2017human}.

Query performance prediction uses query features that are extracted prior to running the query through the system (pre-retrieval) and/or from the initially-retrieved documents (post-retrieval). Intuitively, a good QPP feature should significantly correlate with the actual effectiveness of the IR system. 
Post-retrieval QPP features have been found to be more effective that pre-retrieval features, although they 
are much more expensive to calculate, as they need the IR system to run the query in order to make the prediction.
While the first studies on QPP used single features~\cite{CronenTownsend2002,mothe2005,hauff2009combining,shtok2010using,cummins2014document}, a more recent path is to combine various query features~\cite{zhou2007query,hauff2009combining,raiber2014query,roitman2018extended,vicente2018predicting}. While combining multiple post-retrieval features improves accuracy, the method becomes applicable in real-world scenarios only if the number of features is limited to just a few, due to the increased computational time required for obtaining these features. 
A state-of-the-art method for combining QPP features, that, however, does not take into account these critical issues, is 
Raiber et al.'s~\cite{raiber2014query}. Their system uses a pairwise learning-to-rank model 
that combines several existing 
QPP features. 
It uses a large number of features, deeming it unlikely to be implemented in real-world systems. Moreover, their method results in a non-interpretable model, due to the employed machine learning (ML) method. 

 \emph{Model interpretability} refers to fairness, accountability, and transparency in machine learning~\cite{adler2018auditing,castillo2019fairness,lepri2018fair}, either for compulsory reasons (e.g. in the banking domain, the decision on accepting/rejecting mortgage) or because the end users want to understand the decisions taken by the ML model~\cite{zhao2014menos,andreou2018investigating}. 
 Although fairness and transparency are not yet considered as requirements by search engine users, these features could become more popular with the growing awareness of the influence of search engines on social media users' opinion through the information these engines recommend to the users using sophisticated ML algorithms based on past queries (e.g. influence in political pools, fake news diffusion or unwanted ads). Moreover, public authorities may also require transparency in the near future for users' rights defense and privacy purposes~\cite{song2018personal}. Linear models (e.g. linear regression, SVM with the linear kernel) ensure this transparency, although some recent studies also explain how deep networks make decisions~\cite{chattopadhay2018grad,selvaraju2017grad}.

Nevertheless, to our best knowledge, there is no previous effort to build interpretable models for QPP. In addition to the previously mentioned advantages, an interpretable QPP model would have another huge advantage over non-interpretable ones, considering our lack of understanding of query difficulty. Gaining additional insights from an interpretable model about the difficulty of a query would allow us to propose means for the system to overcome this difficulty. 

In this paper, we propose a QPP approach that combines various features, yet, results in an interpretable and transparent model, so that we know the influence of each feature on the prediction. As a matter of fact, interpretability and transparency should not prevail in detriment of effectiveness. Hence, the model interpretability vs. effectiveness trade-off challenge corresponds to our first research question:
\begin{flushleft}
	\textbf{RQ1}: \textit{
    Can we design an interpretable and transparent model for QPP that is as effective as  complex state-of-the-art black-box models?}
\end{flushleft}

One could argue that the prediction performance improves as the model considers more and more features. However, this statement holds only up to some point~\cite{trunk1979problem}, due to the curse of dimensionality. The curse of dimensionality is particularly problematic when few training examples are available, which happens for QPP evaluated on international reference collections (the only ones available for academic research). A smaller number of features reduces the model uncertainty and improves performance because fewer parameters have to be estimated in the model. Moreover, using more features increases the processing time of the model to the point where it becomes less applicable in real-world scenarios. These considerations are mentioned in present guidelines for IR practitioners~\cite{guyon2003introduction,stapleton2009linear}, being crucial for QPP due to the reliance on post-retrieval features which ensure effectiveness. On the other hand, feature selection poses the challenge of finding the appropriate criteria or strategies to select features in an optimal way. We tackle this problem in our second research question:
\begin{flushleft} 
\textbf{RQ2}: \textit{How selective can a white-box model be, without degrading prediction performance as compared to a non-selective one?}
\end{flushleft}

To solve our two research questions, we develop a new feature selection model for QPP, whose main advantage compared to other related feature selection and QPP models is that it is parameter-free, making it applicable without tuning.

More precisely, our proposed framework is based on an iterative model selection procedure founded on linear regression, one of the most popular yet readable ML approaches. Linear regression is also known for its simplicity, following the Ockham's razor problem-solving principle that essentially states that ``simpler solutions are more likely to be correct than complex ones"~\cite{blumer1987occam,rasmussen2001occam}. 
However, rather than calculating the importance of each feature in one shot as linear regression does, 
we implement an iterative process which, at each step, adds a new feature to test or removes the least performing feature. Moreover, our approach uses a model selection criterion and is able to consider a large set of candidate features. While this approach has been used in machine learning~\cite{claeskens2008model}
, to our knowledge, it has never been considered in QPP. 

In terms of performance prediction, we found that our model is consistently better than non-selective linear regression. We also compare our model with the penalized regression model called LASSO~\cite{hauff2009combining}, which selects features by shrinking some feature coefficients to zero. The results reveal that our proposed model outperforms LASSO in most of the cases. Moreover, compared to LASSO, our method is parameter free and keeps fewer features, thus being less costly to use in the real world. Finally, we compare our model to that of Raiber \textit{et al.}~\cite{raiber2014query}, which is the most recent approach that combines features and uses the same evaluation setting as ours. We found no statistically significant difference between our model and that of Raiber \textit{et al.}~\cite{raiber2014query}, while our model is simpler, interpretable, and uses fewer features. We also investigate the  inference times required by our model versus Raiber’s et al.~\cite{raiber2014query} model, during prediction of the query performance. The time evaluation shows that our model requires less time than Raiber’s et al.~\cite{raiber2014query}. We also found that our method consistently selects a specific group of features across collections for different folds and trials; one of these features is the \emph{QFTERM} proposed in this work.

The remainder of this paper is structured as follows. {Section}~\ref{sec:RW} includes the related work. {Section}~\ref{sec:Model} presents our framework of stepwise model selection for query performance prediction. {Section}~\ref{sec:Eval} presents the data collections, evaluation metrics, and experimental settings used for the evaluation part. {Section}~\ref{sec:RQ} reports the evaluation of the proposed framework and the answers to our research questions. 
Finally, {Section}~\ref{sec:Conclusion} concludes this work and presents some future directions.

\section{Related Work}
\label{sec:RW}

The core objective of our paper is to define an optimal readable model that combines query features, selects the most important ones, and can explain the predicted values as opposed to black-boxes. Work related to our paper is about (a) query performance predictors and (b) methods to combine predictors.

\vspace{0.2cm}
\textbf{Query performance predictors.}
Query performance prediction aims at automatically estimating the performance of a query~\cite{Yom-Tov2005} without relevance judgment.
\textit{Pre-retrieval predictors} were defined first, and they can be calculated prior to any search for the given query. 
Examples of pre-retrieval predictors are the Inverse Document Frequency (IDF)~\cite{Sparck-Jones1972} 
or SynSet (the average number of query term senses)~\cite{mothe2007}. Further pre-retrieval QPPs have been defined in the literature, including the \textit{CLARITY score}~\cite{CronenTownsend2002}, 
the \textit{query complexity}~\cite{mothe2005}, 
and the \textit{query scope}~\cite{He2004}. 
However, the post-retrieval features have been shown to be more effective~\cite{hauff2008survey,shtok2010using,DBLP:conf/clef/MolinaMRTU17}.

\textit{Post-retrieval predictors} require to search through the documents to compute their scores and thus to predict the query difficulty. For example, Diaz~\cite{diaz2007performance} found that ``low correlation between scores of topically close documents often implies a poor retrieval performance" and suggested a spatial analysis of retrieval scores for QPP. Indeed, several QPPs from the literature rely on document scores. Examples of post-retrieval predictors are: the agreement between the entire query results and the results obtained when using sub-queries~\cite{Yom-Tov2005}, Query Feedback (QF)~\cite{zhou2007query}, 
Weighted Information Gain (WIG)~\cite{zhou2007query}, CLARITY~\cite{CronenTownsend2002}, 
Normalized Query Commitment (NQC)~\cite{shtok2012predicting}, 
and score-distribution models~\cite{cummins2014document}. 
Roitman \textit{et al.} proposed an enhanced QPP estimator based on calibrating the retrieved document scores through learning document-level features~\cite{roitman2017enhanced}. Zamani \textit{et al.}~\cite{Zamani2018NQP} proposed a NeuralQPP method based on integrating the retrieval scores, the term distribution, and the continuous representation of the top-retrieved documents by training a neural network with multiple weak supervision signals.

\vspace{0.2cm}
\textbf{Combining query features.}
Several previous studies attempted to combine multiple query features or predictors. Bashir~\cite{Bashir2014} employed a genetic algorithm to combine multiple pre-retrieval features and showed that it 
is more effective than using any single predictor. 
However, a straightforward way to combine features to predict a target value is by linear regression, and most of the related works combining features that way
~\cite{zhou2007query,hauff2009combining}.

Zhou and Croft~\cite{zhou2007query} combined WIG and QF post-retrieval predictors in a linear way, showing that the combination improves performance. 
Shtok \textit{et al.}~\cite{shtok2010using} proposed a framework based on statistical decision theory to estimate the utility of a document ranking for QPP, considering four predictors (WIG, QF, CLARITY, NQC). They reached to the same conclusion as~\cite{zhou2007query}, namely that WIG and QF are worth combining.
Collins-Thompson \textit{et al.}~\cite{collins2010predicting} used a regression tree for QPP by combining features based on divergences between language or topic model representations, such as simplified clarity~\cite{he2006query}, query drift~\cite{winaver2007towards}, clarity~\cite{CronenTownsend2002}, expansion drift~\cite{zhou2006ranking}, and expansion clarity~\cite{collins2010predicting}. 

Hauff~\cite{Hauff2010} used the absolute shrinkage and selection operator (LASSO) penalization when combining pre-retrieval features using linear regression. The LASSO penalization in linear regression aims at making the model sparse by removing features that roughly correspond to the smallest coefficients of the model. Even if LASSO exhibits proficiency in selecting the most important features, it relies on a parameter that has to be tuned optimally, for instance, using a cross-validation approach that can be time-consuming. Therefore, we rather opted for a stepwise approach with a straightforward implementation  since it is parameter free.

Another closely related work is that of Raiber \textit{et al.}~\cite{raiber2014query}. They proposed a pairwise learning-to-rank model,  
that combines several existing pre-and post-retrieval QPP features through a two-stage training~\cite{svmrank2006} process. 
In the first stage, the SVM-rank-based training combines several variants of individual post-retrieval features (e.g. NQC~\cite{shtok2012predicting}) calculated for different hyper-parameter values and for several QPP features. In the second stage, another SVM-rank-based training combines all the QPPs from the first stage, while weighting them according to the weights learned in the first step. The main reason why the two stages are needed is the (large) number of features (and/or feature variants) compared to the relatively small number of training examples. Although this framework~\cite{raiber2014query} shows convincing performance, (a) it is computationally expensive at inference time since all the features are used in the final model, for both training and inference. Moreover, there are as many SVM-rank-based training procedures as the number of QPP features (first stage) plus an additional SVM training in the second stage and (b) the method is not parameter free, specifically in its adaption to sparse SVM. 
Since it is computationally expensive to extract the many QPP features required by Raiber's framework~\cite{raiber2014query} in the second stage, we rather develop 
a selective model which requires only a few features.

\section{Stepwise Model Selection for QPP}
\label{sec:Model}

In this section, we describe our novel framework for selecting features to be used in the query performance predictive model. It employs an iterative process which relies on model selection theory in the context of linear regression and aims at combining various query features into a readable model. Not all the features are equally important and our model aims at optimizing the feature selection. Moreover, we use an iterative algorithm in order to select the best predictive model. While our model belongs to the group of models that have a solid mathematical background, we think it is worth providing the basics of linear regression for readers unfamiliar to ML, since our model is based on an adaption of it.

\vspace{0.2cm}
\textbf{Linear model as a basis for predictor combination.}
Our model is founded on the theory of linear models~\cite{rencher2008linear}. A linear model links a response variable $y$ to several predicting variables $x^{j}, j=1,\dots,p$.
In our context, $x^{j}$ refers to a query feature and $y$ refers to a performance measure representing the ground-truth effectiveness, that our model aims at predicting. We can model the performance measure according to query features and express it as the following linear model:
\begin{equation}
y_i = \beta_1 \cdot x^1_i + \beta_2 \cdot x^2_i + \dots + \beta_p \cdot x^p_i + \beta_0 + \varepsilon_i, \forall i \in \{1,\dots,n\},
\end{equation}

where $i$ is the index of a query. A standard assumption for linear models is that $\varepsilon_i \backsim \mathcal{N}(\mu,\sigma^2)$, expressing that the residuals contain only noise.
Equivalently, the linear model can be expressed using vector and matrix notations:
\begin{equation}
Y = X \beta' + \varepsilon,
\end{equation}
where $Y$ and $\varepsilon$ are $n$-dimensional vectors, $\beta$ is a $(p+1)$-dimensional row vector of weights, $\beta'$ is the transposed (column) vector, and $X$ is a $n \times (p+1)$ matrix containing $n$ training examples.
Our choice toward a linear model for QPP is driven by better interpretability and by the theoretical background we can rely on.

\vspace{0.2cm}
\textbf{Parameter estimation.}
In the linear model, the unknown parameters $\beta_j$ can be estimated using maximum likelihood. In statistics, the likelihood function expresses the way the parameters to be estimated are associated to the data actually observed. Maximizing the likelihood function consists in finding the values of the parameters that plausibly describe the observations. Using the previously defined setting, the likelihood function is given by:
\begin{equation}
L(\beta, \sigma^2) = {(2\pi\sigma^2)}^{-n/2} \mbox{exp} \left( {-\frac{1}{2\sigma^2}\sum_{i=1}^{n}(y_i-X_i\beta)^2} \right).
\end{equation}
Maximizing $L(\beta, \sigma^2)$ is equivalent to maximizing $\log L(\beta, \sigma^2)$ which is easier to handle and maximize. We thus want to maximize:
\begin{equation}
\log L(\beta, \sigma^2) = -\frac{n}{2} \log(2\pi) - \frac{n}{2} \log(\sigma^2) - \frac{1}{2\sigma^2} \sum_{i=1}^{n}(y_i-X_i\beta)^2.
\end{equation}
The maximum (log-)likelihood is reached when the partial derivative according to each parameter is zero. This leads to the following estimators for $\beta$ and $\sigma^2$:
\begin{equation}
\begin{cases}
{\hat \beta} = (X'X)^{-1}X'Y
\\
{\hat \sigma^2} = \frac{1}{n}(Y-X{\hat \beta})'(Y-X{\hat \beta})
\end{cases}
\end{equation}

Once the parameters are estimated, the model can infer the fitted values for the performance measure $y$. Statistical testing (or equivalently confidence intervals) can be used to assess the significance of the parameters. The null hypothesis relies on the nullity of the parameters, i.e. on the uselessness of the associated features to predict the effectiveness of the system. It is interpreted through one p-value associated with each parameter. 
The p-value can be viewed as the probability to make an error when rejecting the null hypothesis. In other words, it is the probability of considering that the feature is not relevant to predict the performance measure, while the opposite is true. Based on the p-value, a feature could be excluded but that feature might be important if combined with others. Thus, we consider combining features rather than accepting or rejecting individual features.

Another crucial issue in linear modeling is variable selection, especially when dealing with a relatively large number of predicting variables. To address this issue, one has to go beyond elementary indicators, that mechanically increase with the number of variables. We chose to focus on the Akaike Information Criterion (AIC)~\cite{claeskens2008model} for assessing the goodness of fit of a linear model.

\vspace{0.2cm}
\textbf{Akaike Information Criterion.}
This criterion is defined from the log-likelihood and uses a penalty to limit the number of parameters in the model. 
The function to minimize is defined as follows:
\begin{equation}
AIC = -2 \log L(\beta, \sigma^2) + 2\cdot k, 
\end{equation}
where $k$ is the number of retained predictors.

AIC aims at selecting the most important features of the linear model. If a feature is kept, then a parameter is estimated to assess its influence in the linear model.
As mentioned above, the parameters in a linear model can be estimated through the maximization of the (log-)likelihood. AIC is based on the opposite of the log-likelihood, thus requiring minimization. However, the penalty term added in AIC ($2\cdot k$) depends on the number of parameters $k$ to be estimated in the model (the same as the number of features included to predict the performance): the higher the number of parameters, the higher the penalty. Therefore, using AIC will ensure that the model does not use ``too many" features and will keep only the most significant ones. Moreover, AIC can be used as a stopping rule for stepwise algorithms for model selection, as shown below.

\vspace{0.2cm}
\textbf{An iterative stepwise selection algorithm.}
Our proposed framework is based on a model selection approach. This means different models are fitted with different subsets of features and the system selects the best model. This selection process is iterative. 

More precisely, we employ a stepwise algorithm which mixes two strategies: forward and backward. Basically, the forward strategy starts from the model with no predictors and adds at each step the feature with the smallest p-value, thus possibly, the most useful because its coefficient in the linear model can be considered as significantly different from zero with a very low risk (quantified by the p-value) to be wrong. A stopping rule is based on a threshold for the p-value. On the other hand, the backward strategy starts from the complete model with all available features and, at each step, removes the feature with the highest p-value. In this case, the p-value can be interpreted as the probability to make
an error if we consider that the coefficient of the feature is not null. The stepwise strategy combines the forward and backward strategies by attempting to remove a feature (applying backward) each time another one is added in the model (applying forward). This strategy is improved using AIC 
as a criterion instead of considering a threshold on the p-value. That is what we use in the following. Although this approach has been used in other ML tasks, it has never been used for QPP. We believe that AIC is worth investigating because of the cost of using multiple post-retrieval prediction features.

\begin{figure*}[!t]
\centering
\includegraphics[width=0.95\textwidth]{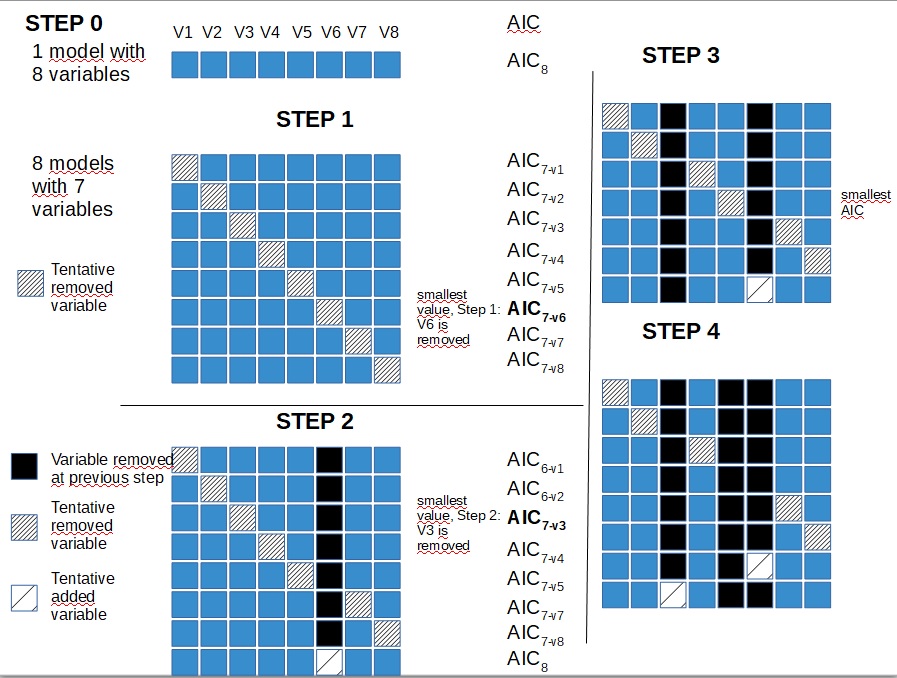}
\vspace*{-6.5pt}
\caption{The four first steps of AIC stepwise model selection when starting with 8 variables. At each step, the best model is kept, either from removing a variable or adding one. }
\label{fig:AIC}
\end{figure*}

Our stepwise algorithm is an automatic model specification based on the AIC criterion. 
Starting from the complete model, 
the stepwise algorithm aims at decreasing the AIC 
at each step, using one of the two possible operations: (a) Remove one variable (obviously, this is the only option at the first step when starting from the complete model); (b) Add one variable removed in an earlier step.

The algorithm stops when the AIC criterion cannot be further decreased by removing or adding a variable. We note that our iterative feature selection algorithm is employed only at train time and it does not affect inference time.

Figure~\ref{fig:AIC} illustrates the selection process when eight variables are used. The initial model consists of 8 variables. In Step 1, we cannot add any variable, the only possibility is to remove one. Eight models are built consisting each of $7$ variables. Let us assume that the model without $V_6$ got the lowest AIC. The model without $V_6$ is the starting point for Step 2. In Step 2, we can either add $V_6$ or remove one of the other 7 variables. We thus test these 8 possible models and keep the one with the lowest AIC. Let us assume that removing $V_3$ is the best. We now have a model with $6$ variables where $V_3$ and $V_6$ do not belong to. This model is used to start Step 3. We can either add one of these 2 variables or remove the third one (6 possibilities). Again, we test all the models. Let us assume that removing $V_5$ leads to the smallest AIC; we would keep the model with $5$ variables for the next step.

\section{Data Collections and Evaluation}
\label{sec:Eval}

\textbf{Data collections.}
We considered four standard TREC collections from the ad-hoc task as follows: Robust, GOV2, WT10G, and ClueWeb12-B13. For Robust, there are approximately 500K newspaper articles. 
WT10G is composed of 1.6 million web/blog page documents. GOV2 includes 25 million web pages and
ClueWeb12-B13 subset 
includes 50 million web pages. Table~\ref{tab:collection} summarizes a few features about the collections used for evaluation. The four TREC test collections also include topics. The ``standard" format of a TREC topic statement comprises a topic ID, a title, a description, and a narrative. 
In our experiments, a query is composed of the topic title that contains two or three words 
representing the keywords a user could have used as a search query. 
Finally, the collections provide \textit{qrels} (i.e. judged documents, relevant or non-relevant, for each query), which are used by the evaluation program \textit{trec\_eval}\footnote{http://trec.nist.gov/trec\_eval/} in order to calculate the effectiveness of the IR system.

\begin{table}[!t]
\setlength\tabcolsep{3.0pt}
\vspace{-0.4cm}
\caption{Details of the collections used in the experiments.}
\vspace{-0.4cm}
\begin{center}
\begin{tabular}{l|r|r}
\textbf{Corpus} & \#\textbf{Docs} & \textbf{Queries} (Title only)\\
\hline
Robust & 528,918 & 301-450, 601-700\\
WT10G & 1,692,096 & 451-550\\
GOV2 & 25,205,179 & 701-850\\
Clueweb12B (CW12B) & 52,343,021 & 201-300\\
\end{tabular}
\label{tab:collection}
\vspace{-0.5cm}
\end{center}
\end{table} 

\vspace{0.2cm}
\textbf{Query performance predictors.}  
Several post-retrieval features have been proposed in the literature as QPP features and we reuse the main ones in this paper. We also propose two new post-retrieval features named \emph{QFTERM} and \emph{QFJSD} as variants of \emph{QFDOC}~\cite{zhou2007query}. Our proposed QPP features, as well as the state-of-the-art ones, are described as follows:\\
\noindent
- \textit{QFDOC}~\cite{zhou2007query}: estimates the query feedback as the percentage of overlap at some rank between the returned document lists for the original query and the expanded query induced from the initially retrieved documents. 
It measures the query-drift.
\\- \textit{QFTERM} (ours): we argue that the overlap at the document level, as computed by QFDOC, is 
too strict to estimate the discrepancy. We thus propose to relax this phenomenon at the term level, 
computing the percentage of overlap between the list of terms available in the top-retrieved documents for the original query and the term list for the expanded query. 
The higher the percentage of term overlaps, the higher is the chance that the top-retrieved documents cover many relevant documents, since the expanded query is not too drifted away from the original query.
\\- \textit{QFJSD} (ours): instead of computing the percentage of overlap at the document level between the top-retrieved documents for the original and the expanded queries, as QFDOC does, we rather measure the query feedback based on the similarity of term statistics between the two document lists, considering that a higher similarity value should correspond to a lower query-drift and a higher query performance. To estimate the similarity of term statistics between the two lists, we first build language models from the top-retrieved documents for the original query and the expanded query, respectively. Then, we apply the Jensen-Shannon divergence between the two language models to estimate how similar they are.
\\- \textit{CLARITY}~\cite{CronenTownsend2002}:  estimates the relative entropy between the relevance language models of the top retrieved documents and the corpus.
\\- \textit{WIG}~\cite{zhou2007query}: corresponds to the divergence between the mean of the top-retrieved document scores and the mean of the entire set of document scores.\vspace{.1cm}
\\- \textit{NQC}~\cite{shtok2012predicting}:  is based on the standard deviation of the retrieved document scores.
\\- UQC~\cite{shtok2012predicting}: is a variant of the \emph{NQC} predictor,  based on the standard deviation of the retrieved document scores without normalization.
\\- SW1~\cite{raiber2014query}: is the ratio between the number of stop and non-stop words in each document, averaged over the top-retrieved documents for a query.

These QPP features can be estimated for different numbers of $n$-top-ranked feedback documents where $n$ is a hyper-parameter. In this work, we consider 6 values of $n$ = \{10, 50, 100, 200, 500, and 1000\}. Moreover, to compute \emph{QFDOC}, \emph{QFTERM}, and \emph{QFJSD}, we need to know the cutoff rank (termed QFcut) at which the percentage of overlap is computed for each hyper-parameter $n$. According to common practice~\cite{shtok2012predicting,raiber2014query,roitman2017enhanced}, we define $QFcut = min(50, n)$, i.e. the percentage of overlap is calculated for at most 50 documents.

\vspace{0.2cm}
\textbf{Evaluation metrics.}
We use the Pearson and Spearman correlations between the predicted effectiveness value and the ground-truth effectiveness, as in previous works~\cite{hauff2009combining,carmel2010estimating,roitman2017enhanced,Chifu2018,Zamani2018NQP,roitman2018extended}. To measure the ground-truth effectiveness of the system, we use AP (average precision) and NDCG (normalized discounted cumulative gain) since they are commonly adopted in related works~\cite{hauff2009combining,raiber2014query,roitman2017enhanced,Chifu2018,Zamani2018NQP}.

\vspace{0.2cm}
\textbf{Experimental settings.}
As a common practice in QPP evaluation~\cite{raiber2014query,roitman2018extended,Zamani2018NQP}, 
we randomly split the queries into two equally-sized sets and conduct two-fold cross-validation. We repeat these steps for 30 times and report the average results. Statistically significant differences of prediction performance are estimated using two-tailed paired t-test with Bonferroni correction (p $<$ 0.05) computed over the 30 splits. 
Similar to previous works~\cite{CronenTownsend2002, zhou2007query, shtok2010using, raiber2014query}, we chose the Language Modeling with Dirichlet smoothing and $\mu=1000$ without query expansion (as implemented in Lemur Indri platform, using default parameters) to retrieve $n$ documents for each query and to calculate the performance of the IR system (and thus, determine the results to be predicted in terms of AP or NDCG).

\section{Results and Discussions}
\label{sec:RQ}
\begin{table*}[t]
\setlength\tabcolsep{0.2pt}
\caption{Correlation of predicted and ground-truth effectiveness for scores with Pearson ($r$) and for ranks with Spearman ($\rho$). The first line uses a one step process. The others use a two steps process, the first step being SVM-rank based. $\vartriangle$  indicates statistically significant improvement  over the Linear model (1S LM). $\uparrow$ (resp. $\downarrow$) indicates statistically significant increase (resp. decrease) over the model of Raiber et al.~\cite{raiber2014query}, according to a paired t-test ($p < 0.05$) with a Bonferroni correction. 
\vspace{-0.3cm}}
\begin{center}
\label{tab:FrameworkEvaluation}
\begin{tabular}{@{\hspace*{0.1cm}}l@{\hspace*{0.1cm}}|c@{\hspace*{0.1cm}}c|c@{\hspace*{0.05cm}}c|c@{\hspace*{0.05cm}}c|c@{\hspace*{0.05cm}}c|c@{\hspace*{0.05cm}}c|c@{\hspace*{0.05cm}}c|c@{\hspace*{0.05cm}}c|c@{\hspace*{0.05cm}}c}
& \multicolumn{4}{c|}{\textbf{ROBUST}} & \multicolumn{4}{c|}{\textbf{WT10G}} & \multicolumn{4}{c|}{\textbf{GOV2}} & \multicolumn{4}{c}{\textbf{CW12B}}\\
\cline{2-17}
\bf Method & \multicolumn{2}{c}{\textbf{AP}} & \multicolumn{2}{c|}{\textbf{NDCG}} & \multicolumn{2}{c}{\textbf{AP}} & \multicolumn{2}{c|}{\textbf{NDCG}} & \multicolumn{2}{c}{\textbf{AP}} & \multicolumn{2}{c|}{\textbf{NDCG}} & \multicolumn{2}{c}{\textbf{AP}} & \multicolumn{2}{c}{\textbf{NDCG}} \\
\cline{2-17}
& $r$ & $\rho$ & $r$ & $\rho$ & $r$ & $\rho$ & $r$ & $\rho$ & $r$ & $\rho$ & $r$ & $\rho$ & $r$ & $\rho$ & $r$ & $\rho$\\
\hline
1S LM (One stage linear model)
& .39$^{~~}$ & .41$^{~~}$ & .41$^{~~}$ & .43$^{~~}$ & .01$^{~~}$ & .01$^{~~}$ & .04$^{~~}$ & .05$^{~~}$ & .11$^{~~}$ & .23$^{~~}$ & .13$^{~~}$ & .25$^{~~}$ & .02$^{~~}$ & .01$^{~~}$ & -.00$^{~~}$ & -.02$^{~~}$\\

SVM of Raiber et al.~\cite{raiber2014query} 
& .48$^{\vartriangle}$ & .50$^{\vartriangle}$ & .49$^{\vartriangle}$ & .49$^{\vartriangle}$ 
& .27$^{\vartriangle}$ & .22$^{\vartriangle}$ & .27$^{\vartriangle}$ & .29$^{\vartriangle}$
& .45$^{\vartriangle}$ & .46$^{\vartriangle}$ & .45$^{\vartriangle}$ & .46$^{\vartriangle}$ 
& .41$^{\vartriangle}$ & .36$^{\vartriangle}$ & .37$^{\vartriangle}$ & .36$^{\vartriangle}$\\[2pt]

LM (Linear model) 
& .46$^{\vartriangle}_{\downarrow}$ & .47$^{\vartriangle}_{\downarrow}$ & .48$^{\vartriangle}_{\downarrow}$ & .48$^{\vartriangle}_{\downarrow}$
& .29$^{\vartriangle}$ & .29$^{\vartriangle}_{\uparrow}$ & .30$^{\vartriangle}$ & .36$^{\vartriangle}_{\uparrow}$
& .47$^{\vartriangle}_{\uparrow}$ & .48$^{\vartriangle}_{\uparrow}$ & .44$^{\vartriangle}$ & .47$^{\vartriangle}$ 
& .40$^{\vartriangle}$ & .37$^{\vartriangle}$ & .37$^{\vartriangle}$ & .34$^{\vartriangle}$\\[2pt]

LASSO 
& .48$^{\vartriangle}$ & .48$^{\vartriangle}$ & .49$^{\vartriangle}$ & .49$^{\vartriangle}$ 
& .21$^{\vartriangle}$ & .20$^{\vartriangle}$ & .23$^{\vartriangle}$ & .27$^{\vartriangle}$ 
& .48$^{\vartriangle}_{\uparrow}$ & .50$^{\vartriangle}_{\uparrow}$ & .45$^{\vartriangle}$ & .48$^{\vartriangle}$ 
& .38$^{\vartriangle}$ & .35$^{\vartriangle}$ & .32$^{\vartriangle}$ & .30$^{\vartriangle}_{\downarrow}$\\[2pt]
AIC FS (Feature selection) 
& .46$^{\vartriangle}$ & .47$^{\vartriangle}_{\downarrow}$ & .49$^{\vartriangle}$ & .49$^{\vartriangle}$ 
& .24$^{\vartriangle}$ & .24$^{\vartriangle}$ & .25$^{\vartriangle}$ & .29$^{\vartriangle}$ 
& .45$^{\vartriangle}$ & .47$^{\vartriangle}$ & .43$^{\vartriangle}$ & .45$^{\vartriangle}$ 
& .38$^{\vartriangle}$ & .35$^{\vartriangle}$ & .33$^{\vartriangle}$ & .30$^{\vartriangle}_{\downarrow}$\\

\end{tabular}
\end{center}
\end{table*}

\vspace{0.3cm}
\textbf{Trade-off between sparsity and effectiveness.}
To answer our two research questions, we first study the correlation between the predicted and the ground-truth effectiveness (Table~\ref{tab:FrameworkEvaluation}). When the Pearson correlation coefficient is employed, we measured the correlation between the predicted value and the actual value (the reference system is the language model with $\mu=1000$). When the Spearman correlation coefficient is used, we measured the correlation between the ranks of the queries obtained when ordered by the predicted effectiveness and the actual effectiveness. 


In Table~\ref{tab:FrameworkEvaluation}, the first row is obtained by using the linear model (LM) with all the features in a single step (1S), a baseline that achieves readability, but not sparsity. The models listed on the subsequent rows use a two-stage approach as in~\cite{raiber2014query}, where the second step is either SVM-rank~\cite{raiber2014query} (second row) or one of the models that ensure interpretability, as follows: 
LM (third row) refers to the linear model; LASSO (fourth row) is the LASSO selection~\cite{lasso} that ensure sparsity. 
Finally, the last row (AIC FS) corresponds to our proposed method of feature selection (FS) using the AIC criterion. 

From the results displayed in Table~\ref{tab:FrameworkEvaluation}, we observe that Raiber et al.'s~\cite{raiber2014query} two-stage framework outperforms the one-stage linear regression baseline. Indeed, rows 2 to 5 indicate significant increases in correlation compared to the first baseline (see $\vartriangle$ in the table), irrespective of the model, the collection or the correlation measure being used. This result was expected considering the state-of-the-art results, but was worth checking\footnote{The linear model with a single step (1S LM) does not work well apart from the Robust collection.}. 

More interestingly, we notice that, in Table~\ref{tab:FrameworkEvaluation}, the three models we implemented (LM, LASSO and AIC FS) are generally (a) close to one another in terms of results, (b) without significant differences with respect to Raiber et al.~\cite{raiber2014query}, apart from a few cases (see $\downarrow$ and $\uparrow$ in the table). These results show that it is possible to use an interpretable model without decreasing effectiveness. From these three models, only LASSO and AIC FS ensure sparsity; we thus focus next on the results obtained by these models.

LASSO and AIC FS perform almost the same apart from a few cases where LASSO is slightly better (ROBUST AP and GOV2) or, conversely, where AIC is slightly better (WT10G). However, one important observation is that LASSO has a shrinkage parameter $\lambda$ that needs to be tuned. In our experiments, we fitted it using 10-fold cross-validation, thus using the same data for both parameter fitting and model training, giving a clear advantage to LASSO in our experiments. 
In a preliminary set of experiments, we found that the transfer learning of $\lambda$ (learning the parameter on a collection and using that value for another collection), that would make a fair comparison between LASSO and AIC FS, did not work at all for LASSO. Hence, $\lambda$ has to be tuned separately for each collection, which is a clear drawback of LASSO compared to our parameter-free selection model based on the AIC criterion.

While one could have hypothesized that using all the features (the 1st and 3rd rows in Table~\ref{tab:FrameworkEvaluation}) would have outperformed all the selective methods, this is not the case. This result is likely due to the curse of dimensionality~\cite{trunk1979problem}. This is an important result, as one crucial advantage of selective methods is to avoid calculating some features at inference time, that are not kept in the trained model. This also follows the Ockham's principle ``the law of briefness", that is ``more things should not be used than are necessary." In the case of AIC, not only a limited number of features have to be computed, but the results are also better than using all features.

\vspace{0.2cm}
\textbf{Sparsity and time complexity of the resulting models.}
As eight predictors 
are not that many, one may argue that this is a reasonable number which does not require feature selection. However, each feature can take valuable time in order to be computed, e.g. WIG requires 3.8 seconds on average for each query from the TREC-ROBUST collection on a machine having 8GB of RAM and processing on a single core. 
Thus, sparsity is an important issue when the number of predictors is large and/or costly to compute. Moreover, a smaller number of predictors leads to a simpler model and easier interpretation~\cite{zou2005regularization}. 

To investigate the sparsity of our model, we computed the number of features selected in the second stage for the different models. We start from eight QPP features in the second stage, two folds, and 30 trials. 
Since different numbers of features may be selected by a model across trials, we compute the average number of selected features. 
Table~\ref{tab:Nb} reports the average number of features selected by Raiber et al.~\cite{raiber2014query} and AIC FS models. 
Remarkably, our method, AIC, 
selects much fewer features than Raiber et al.~\cite{raiber2014query} 
across all collections. The average number of features selected by AIC ranges from $2$ to $5$, while maintaining a similar performance to SVM-rank~\cite{raiber2014query}, that uses $7$ to $8$ features. This makes our method more applicable in real-world systems. 

In Table~\ref{tab:Nb}, we also report the inference times required by our model versus Raiber's et al.~\cite{raiber2014query} model, respectively, during prediction of the query performance. The time evaluation shows that AIC FS takes less time than Raiber's et al.~\cite{raiber2014query} SVM to predict the query performance, since AIC FS requires fewer QPP features.

\vspace{.2cm}
\begin{table}[!t]
\setlength\tabcolsep{2.5pt}
\centering
\caption{Average number of features selected in the models and average inference time 
required to predict the query performance using the same configurations as Table~\ref{tab:FrameworkEvaluation}. We use a machine with an AMD Opteron 6262HE 1.6 GHz CPU, 8GB of RAM, a single thread.\vspace{-.4cm}}
\label{tab:Nb}
\begin{tabular}{l|l|c|c|c|c}
& Method & \multicolumn{1}{c|}{\textbf{ROBUST}} & \multicolumn{1}{c|}{\textbf{WT10G}} & \multicolumn{1}{c|}{\textbf{GOV2}} & \multicolumn{1}{c|}{\textbf{CW12B}}\\
\hline
\multirow{3}{1cm}{AP}
& \textit{Raiber's SVM} & 8 (.042s) & 6 (.036s) & 8 (.033s) & 6 (.023s)\\
& AIC FS & 3 (.019s) & 3 (.025s) & 4 (.023s) & 3 (.015s)\\
\hline
\multirow{3}{1cm}{NDCG} 
& \textit{Raiber's SVM} & 8 (.041s) & 7 (.037s) & 8 (.017s) & 7 (.020s)\\
& AIC FS & 3 (.026s) & 4 (.029s) & 4 (.014s) & 3 (.013s)\\
\end{tabular}
\end{table}

\begin{figure}[!ht]
\centering
\includegraphics[width = 0.9 \linewidth]{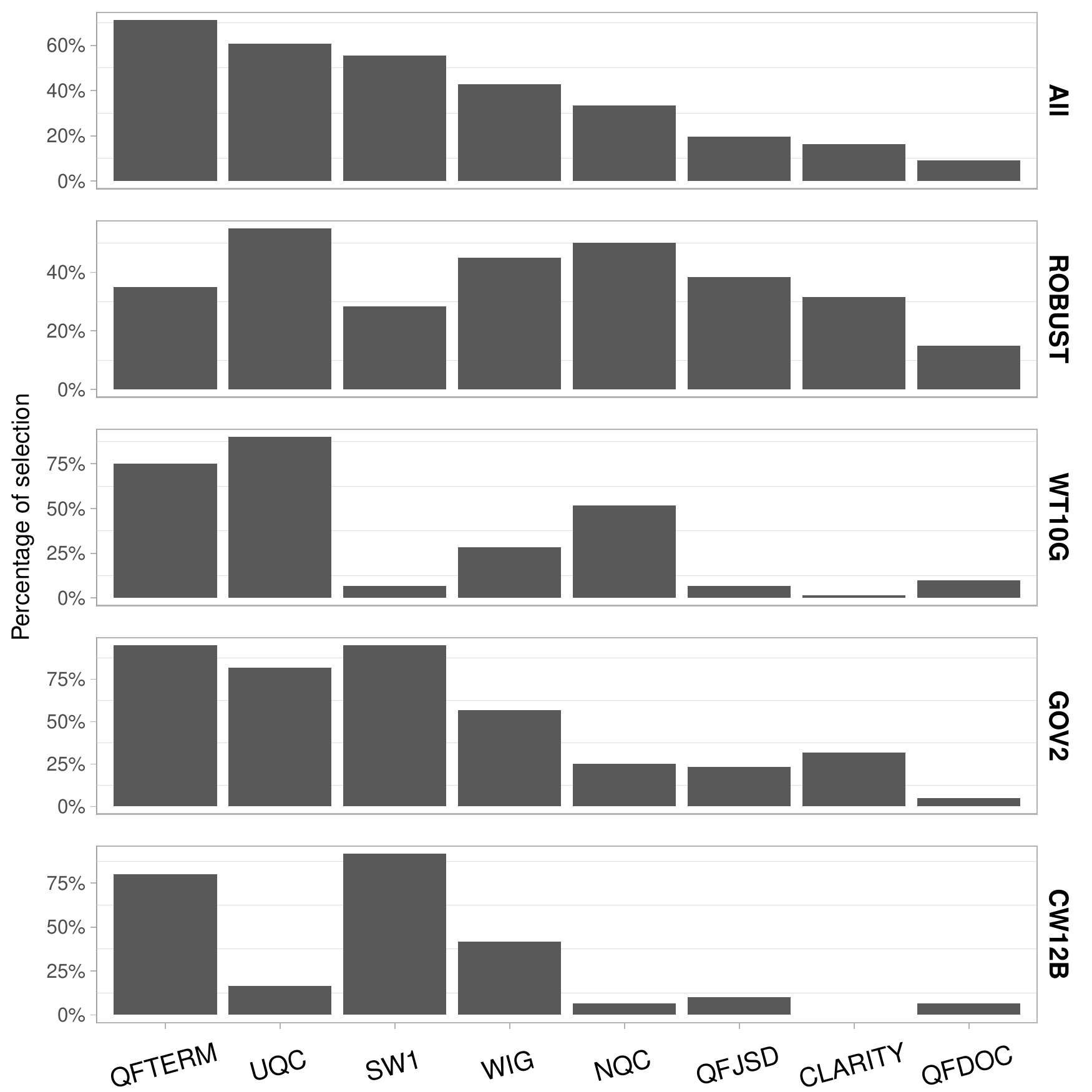}
    \caption{Distribution of features selected by our AIC FS model across collections for AP. 
Each percentage is computed as the number of times a feature is selected divided by the total number of times a feature could be selected.}
\label{fig:fs-ndcg-aic}
\end{figure}


\vspace{0.2cm}
\textbf{Most important features and interpretability.}
For a deeper understanding, we analyze the features selected by our AIC model, presenting the usefulness of each feature in Figure~\ref{fig:fs-ndcg-aic} (similar results are obtained for NDCG). 
We compute the number of times a feature was selected ($SF$) by our model, for two folds in 30 trials. Then, we compute the percentage of selecting that feature as $\frac{SF}{TF}\times 100$, where $TF$ denotes the number of times a feature could be selected, i.e. the number of folds times the number of trials. This is done for all individual features per collection. 

One interesting finding is that \emph{QFTERM}, one of the  feature we proposed in this paper, is consistently selected as an important feature. 
WIG is also consistently important. UQC and SW1 are more collection dependant, the first one being more important for ROBUST, WT10G, and GOV2, while the second one being more important for GOV2 and CW12B. 

The proposed AIC FS method is easily interpretable and enables one to understand the trained model and to know the impact of each query features in QPP. To illustrate this, we include below a model obtained on GOV2 collection: \\
\begin{equation}
\begin{split}
QPP &= .136 \cdot SW1 + .126  \cdot QFTERM + .101 \cdot UQC \\
& + .091 \cdot WIG + .034 \cdot CLARITY + .184 
\end{split}
\end{equation}


We can see that the weight of the two most important features (SW1 and QFTERM) are 30\% higher than the two following ones (UQC and WIG); CLARITY is much less important.


\section{Conclusion}
\label{sec:Conclusion}


In this paper, we promoted the use of model selection for combining query performance predictors. Our aim was twofold: (i) we wanted to contribute to shifting the effort from complex predictive models to simpler models and (ii) we wanted to develop a sparse and easy to interpret 
model.
Indeed, we have shown that the trade-off between simplicity and effectiveness is in favour of our model.
We showed that our selective framework achieves similar results in terms of correlation measures, while having the great advantage of using a limited number of features, and thus, being much cheaper for implementation in real-world systems. Our predictive approach based on the AIC selection strategy provides the best trade-off between the prediction accuracy and the number of features to be computed. 
During inference, the features selected by our framework are the only ones which need to be calculated. That gives a key advantage to our framework compared to the literature. Moreover, our model is interpretable, which is very important at this stage of query difficulty research. These results open the path to a better understanding of system failures by analyzing the model deeper. 

To answer this challenge, in future work, we will analyze the influence of each of the selected features individually as well as their cost/effectiveness trade-off. Because the current results reveal that our new version of QFDOC feature, namely {QFTERM}, is the most important QPP feature in the model across collections and performance metrics, it will be worth to continue defining new QPP features to try to further improve query difficulty prediction. \bibliographystyle{unsrt}
\balance
\bibliography{references} 
\end{document}